\documentstyle[12pt]{article}
\begin{document}

\parskip 2mm plus 1mm \parindent=0pt
\def\cl{\centerline}\def\lel{\leftline}\def\rl{\rightline}
 \def\hs1{\hskip1mm} \def\h10{\hskip10mm} \def\hx{\h10\hbox}
\def\vs{\vskip3mm} \def\vup{\vskip-2mm}
\def\page{\vfill\eject}
\def\<{\langle} \def\>{\rangle} \def\br{\bf\rm}  \def\it{\tenit}
 \def\de{\partial} \def\Tr{{\rm Tr}} \def\dag{^\dagger} 
\def\half{{\scriptstyle{1\over 2}}}
\def\ne{=\hskip-3.3mm /\hskip3.3mm}\def\tr{{\rm tr}}
\def\Pr{{\br Pr}}\def\to{\rightarrow}
\def\vb{\vskip20mm}\def\vm{\vskip10mm}
\def\cite{}

\def\be{\begin{equation}}\def\ee{\end{equation}}

\vs\bf\cl{Why do Bell experiments?}

\vs

\centerline {by}

\vs

\centerline {Ian C. Percival}\vs 

\centerline {Department of Physics} 
\centerline {Queen Mary and Westfield College, University of London} 
\centerline {Mile End Road, London E1 4NS, England} 

\vs\vs\vs

\rm

Experiments over three decades have been unable to demonstrate weak
nonlocality in the sense of Bell unambiguously, without loopholes.
The last important loophole remaining is the detection
loophole$^{1,2,3}$, which is being tackled by at least three
experimental groups$^{4,5,6}$.  This letter counters five
common beliefs about Bell experiments, shows the importance of
these experiments, and presents alternative scenarios for future
developments.
Figure here.

The figure shows the bare bones of a system designed to test Bell
inequalities$^7$.  It may be considered as a black box at rest in a
laboratory frame, with two input ports $\alpha$, $\beta$ and two
output ports $a$, $b$.  The input port $\alpha$ and the output port
$a$ are parts of a subsystem $A$ on the left of the box, and $\beta$,
$b$ are parts of a subsystem $B$ on the right. The minimum distance
between $A$ and $B$ is $L$.

One run of an experiment on the black box starts with inputs labelled
$\alpha$, $\beta$ which end before time $t=0$ in the laboratory frame.
The outputs are labelled $a$, $b$ start after time $t=T$.  The inputs
and outputs are entirely classical, but the system has entangled
quantum components.

A sufficient number of runs provides transition or conditional
probabilities denoted

\be\label{} \Pr(\alpha,\beta\to a,b)  \ee 

for the outputs $a,b$ of the whole system, given the inputs
$\alpha,\beta$.  Following Wigner's formulation$^{8}$, locality
implies that they satisfy the Bell inequality

\be\label{} \Pr(1,2\to +,-) +\Pr(2,0\to +,-) -\Pr(0,1\to
+,-) \ge 0, \ee 

together with two more inequalities given by cyclic
permutations of $0,1,2$.  

The subsystems $A$ and $B$ are so far apart that 

\be\label{} L > Tc, \ee 

where $c$ is the velocity of light, so it is not possible to send
signals from $\alpha$ to $b$ or from $\beta$ to $a$.  Without this
condition, the experiment has a loophole, sometimes called the
locality loophole$^9$.  All early experiments had this loophole, but
Aspect's group$^{10}$ and more recently Zeilinger's and
Gisin's groups$^{11,12}$ have removed it.

In Bell's original thought experiment, there are three possible values
of $\alpha$ and corresponding values of $\beta$, denoted $0,1,2$.
They might be three settings of the angle of a polarizer, or of the
angle of measurement of a particle of spin one-half.  There are two
possible values of each separate output, denoted $+$ and $-$.  They are
typically two orthogonal directions of polarization of a photon, or
two opposite spins of an atom.  In an ideal Bell experiment, the
inputs $\alpha$ and $\beta$ are determined by completely independent
random variables.  If they are not random, as for the Aspect and Gisin
experiments, there is a further loophole, which was removed by
Zeilinger's group in an experiment in which the locality loophole was
also overcome$^{11}$.

The detection loophole$^{1,2,3}$ described below has proved to be
the most difficult of all, and has never been overcome in any
published experiment, though the groups of Fry and Walther$^4$, of
Wineland$^{5}$ and of Blatt$^{6}$ are trying.  Fry and Walther aim
to close the locality loophole in the same experiment.

Bell experiments, or experiments of the Bell type, are experiments to
test the original Bell inequalities, or one of the many other
inequalities that follow from locality.  Here are five common beliefs
about them.

1 Their only purpose is to exclude local hidden variable theories,
which are of little interest anyway.

\vup 2 Violation of the inequalities follows inevitably from the laws of
quantum mechanics and their interpretation.

\vup 3  Experiments have already shown that the Bell inequalities are
violated, apart from the detection loophole, which is so unbelievable
that it is not worth considering seriously.

\vup 4 Einstein's view that all physical laws are local was his one
definite major mistake.  

\vup 5 Bell experiments are therefore no longer worth doing.

Reasons are presented here for rejecting all of these beliefs, and
replacing them by the alternatives:

1 There are more important reasons for doing Bell experiments, including
Bell's weak nonlocality. 

\vup 2 Neither the violation nor the nonlocality follow inevitably from
quantum mechanics.

\vup 3 There are at least two good reasons why the detection loophole
should be taken seriously.

\vup 4 Einstein has been right before, when many in the physics community
were wrong, and we need conclusive experimental evidence of
nonlocality before judging him on {\it this} issue.

\vup 5 Bell experiments are among the most important in physics.

\vs{\bf Why do Bell experiments?} 

Guinness, Bass and Worthington are brands of beer.  It is questionable
whether, if Guinness is good for you, Bass is bad and Worthington is
worse.  These are matters of taste and prejudice.  Forward, nonlocal
and backward causality are brands of causality.  If forward causality
is good for you, nonlocal causality is bad and backward causality is
worse.  These are matters of experience.  We are particularly
concerned with the question of nonlocal causality, in which cause and
effect are spatially separated in spacetime, so that a signal from
cause to effect would have to go faster than the velocity of light.
According to classical special relativity, an event can affect a
future event, in or on its forward light cone, but not a spatially
separated event, and certainly not a past event.
  
But apparently, according to quantum theory, classical events that are
linked by quantum systems are different.  For them, there is a sense
in which causality might act nonlocally, but without any signalling
faster than light.  This is Bell's weak nonlocality, which can be
formulated in terms of the classical inputs and outputs of a black
box $^{13}$.

An electrical engineer's black box consists of a circuit with input
and output terminals.  He may not know what circuit is inside, but it
is assumed here to be classical.  If there is no noise in the circuit,
then the black box is deterministic.  The outputs then depend on the
inputs through a unique transfer function $F$, and by experimenting
with different inputs and looking at the outputs, engineers can find
$F$.  In practice the resistors in the circuit produce noise, which we
assume to be classical noise.  The system is then stochastic.  The
noisy circuit can be represented by a probability distribution
$\Pr(F)$ over the transfer functions $F$, in which the unknown values
of supposedly classical background variables, like the coordinates of
thermal electrons, determine the particular $F$ that operates.

A physicist's black box contains an evolving physical system, such
as a classical electrical circuit, or an entangled quantum state with
classical inputs and outputs.  She may not know what physical system
is inside, but by experimenting with different inputs and looking at
the outputs, she can find out something about it.  

For deterministic systems, special relativity distinguishes between
local transfer functions $F$ in which the influence of an input on an
output goes at no more than the velocity of light, and nonlocal
transfer functions $F$, for which the influences can act over
spacelike intervals.  It is possible to determine whether the transfer
function of a system is local or not by experimenting with different
values of the inputs, and observing the outputs.  There is no need to
look inside the black box.  All real classical systems have local
transfer functions, as required by special relativity.

When classical or quantum systems are stochastic, special relativity
distinguishes between three types of black box system, defined in
terms of probabilities $\Pr(F)$ of transfer functions.  The first are
local systems, for which the transition probabilities of the outputs
given the inputs can be obtained from a probability distribution
$\Pr(F)$ in which only local $F$ contribute.  It is therefore not
possible to send signals faster than the velocity of light.  For the
second type, the transition probabilities can only be obtained from
$\Pr(F)$ in which at least one nonlocal transfer function has nonzero
probability, so there is an element of nonlocality.  But nevertheless
it is not possible to send signals faster than the velocity of light.
The system is then said to be weakly nonlocal, or nonlocal in the
sense of Bell.  For the third type, which has never been seen, it is
possible to send signals faster than the velocity of light.

The stochasticity of classical systems comes from background variables
that are not included in the system, but for quantum systems it does
not come from any background variables that we can see, so either they
are assumed not to exist, as in the Copenhagen interpretation, or they
are called hidden variables.

A Bell experiment is a black box with classical terminals and an
entangled quantum system inside, where the source of entanglement is
inside the box.  For photon polarization the setting of the
orientations of the polarizers is an input, and the detection of the
directions of polarization is an output.  All the inputs and outputs
are classical events.

Real laboratory Bell experiments are treated in the next section.  In
this one we treat only ideal Bell thought experiments in which the
entangled quantum system is sufficiently close to a pure state, and
the measurements sufficiently good, that the black box is weakly
nonlocal.  

The classic Bell was proposed to test whether local hidden variable
theories are possible$^7$.  But quantum black boxes also tell us
something about the world: there are correlations between classical
events that can only be produced by quantum links.  These correlations
are important in their own right.  They demonstrate weak nonlocality.
They also show that the properties of our world cannot be explained
using local hidden variables, but that is not their main significance,
which persists independently of any theory, local or nonlocal.
An experimenter who has never seen the apparatus before can
tell by experimenting with the inputs and outputs, and without
looking inside, that the black box contains a quantum system.  This
property of quantum black boxes comes from weak nonlocality.

So Bell experiments and weak nonlocality are important for all quantum
physicists, whether they support hidden variable theories or not.  The
weak nonlocality of quantum measurement is unique in modern physics:
classical dynamics, quantum dynamics and general relativity are all
local.  Today only ideal experiments are weakly nonlocal, though
tomorrow they could be real.

In modern quantum computation it is proposed to put quantum correlations
to good use.  Violation of Bell inequalities is a benchmark experiment
for quantum computers, and for this reason alone would be worth doing
even if there were no interest in weak nonlocality.

\vs{\bf Nonlocality and quantum mechanics}

Real experiments, with or without loopholes, are approximations to
ideal experiments without them.  There are possible limits on
the approach to the ideal that are explored by attempts to carry out
an experiment without loopholes.  Theoretically it appears that it is
possible to approach arbitrarily close to the ideal, by improving the
efficiency of detectors, collimation, etc., but there appears to have
been a `conspiracy' of nature that prevented this.  Such conspiracies
in physics have a long history.

For example, the first law of thermodynamics says that heat is a form
of energy.  In the early 19th century there appeared to be a
conspiracy that prevented anyone from extracting all this energy from
a system and using it.  We now call this conspiracy the second law of
thermodynamics.  Bell's opinion, `It is hard for me to believe that
quantum mechanics works so nicely for inefficient practical setups and
is yet going to fail badly when sufficient refinements have made.'
may be right, but irrelevant.  Quantum theory does not have to fail.
The necessary refinements may not be possible.  Current quantum theory
would then be incomplete, just as the first law of thermodynamics is
incomplete.  Einstein thought that quantum mechanics is
incomplete$^{14}$, but this was not the kind of incompleteness he
described.

Santos has suggested earlier that the laws of quantum measurement
might be compatible with locality $^{15}$.  This idea can be
illuminated by an analogy, comparing the second law of thermodynamics
and the possible breakdown of the nonlocality argument.  Classical
systems obey the laws of Hamiltonian dynamics, despite the second law,
which limits energy transfer from real systems with many
particles.  Similarly quantum systems might obey all the laws of
quantum dynamics and quantum measurement, despite a supplementary law
which excludes weakly nonlocal systems, thus ensuring that physics
remains local.  No one yet knows any such supplementary law.

In thermodynamics, the second law was discovered as a result of many
trials showing practical limitations on getting useful energy from
heat.  Locality holds for all Bell experiments to date, but it is on
much weaker ground, as Bell experiments are relatively few, so
experiments designed to test it are among the most important in
physics.

\vs{\bf Experiments and the detection loophole}

A Bell experiment without loopholes would be an experiment from
which we could deduce weak nonlocality without further assumptions.
It is nearly four decades since the inequalities were obtained and
experiments tentatively suggested, and three decades since the first
experiments.  There is still no published clear experimental
demonstration of weak nonlocality, because of the detection loophole,
which follows.

Real experiments have outcomes that are excluded in ideal experiments.
For example, a photon or an atom may be detected at $A$, but not at
$B$, as a result of imperfect detectors, or losses due to absorption
or bad collimation.  These outcomes affect the inequalities, and the
tests of nonlocality.  There are further assumptions that have to be
made in order to obtain the probabilities that appear in the
inequalities.  One such assumption is that the detector efficiency is
independent of the local `hidden' variables, as discussed by Clauser
and Horne$^{16}$ and by Gisin and Gisin$^{17}$.  If the possibility
of such a dependence is accepted, nonlocality cannot be demonstrated
until the detection efficiency reaches a threshold.  This is the
loophole.

At first sight such a dependence seems unlikely, but there are
two good reasons why the detection loophole should not be ignored.
 
The first reason lies in the definition of efficiency, for which the
analogy with thermodynamics is useful.  If only the first law of
thermodynamics applied, then it would be reasonable to measure the
efficiency of a heat engine as the proportion of the total energy that
is extracted.  But once the second law is recognized, this definition
is inappropriate, and we revise the definition to take temperature
differences into account.  Similarly, if there are values of local hidden
variables that play a role in determining whether or not there is a
response from a particle detector, it is no longer appropriate to
measure the efficiency of the detector in the conventional way.  The
measure of efficiency should take into account the values of the hidden
variables.  With such a new definition, the dependence seems
natural$^{17}$.

The second reason lies in the alternative.  The dependence appears
unlikely to some people, but the alternative is nonlocality, which is
a break with the whole of the rest of physics.  We must not base
conclusions about such an overwhelmingly important universal issue on
some prejudice about the properties of mere detectors.

We cannot dismiss the detection loophole.  We have to try to close it
by improving the experiments.

\vs{\bf Einstein's mistake?}

Einstein introduced light quanta in 1905, but leading physicists still
did not accept them as late as 1913, so we should be careful before
rejecting his other ideas, whatever the majority thinks$^{18}$.
He believed that nature obeys local laws, and Bell showed that this
assumption might be tested experimentally.  It appeared that weak
nonlocality followed from the laws of quantum dynamics and quantum
measurement, but this is not so.

\vs{\bf What now?}

Quantum technology has advanced so much during the last decade that
the detection loophole might soon be closed, using spin states of
atoms or otherwise.  For the future, there are several possible
scenarios, of which two are the most likely:

{\it EITHER}

1. The inequalities cannot be violated.  The apparent conspiracy is
due to a new law of nature, consistent with current quantum theory,
but limiting the accessible states of matter to those for which
locality reigns.  The common view is wrong and Einstein was right.

{\it OR} 

2. The inequalities can be violated.  The apparent conspiracy that has
prevented the unambiguous experimental confirmation of Bell
nonlocality is due to practical difficulties that can be overcome.
Experiments will close the detection loophole.  This would be a
significant advance, a benchmark on the way to quantum computation.
Simultaneous closure of both the detection and locality loopholes
would confirm the common view that the laws of nature are weakly
nonlocal and that Einstein was wrong.

This issue can only be resolved by experiment.  That is why Bell
experiments are so important.

\vs{\bf References}

1. Pearle, P., Hidden-variable example based on data rejection {\it Phys. Rev. D}, {\bf 2} 1418-1425 (1970).

2. Clauser,J.F.,Horne, M.A., Shimony,A. \& Holt, R.A., Proposed experiment to test local hidden variable theories, {\it
 Phys. Rev.  Lett.}, {\bf 23}, 880-884, (1969).

3. Garg, A. \& Mermin, N.D., Detector inefficiencies in the EPR
experiment, {\it Phys. Rev D}, {\bf 35}, 3831-3835 (1987).

4. Fry, E.S. \& Walther,T., Fundamental tests of quantum mechanics,
  {\it Adv. Atom. Molec. and Opt. Phys.}, {\bf 42} 1-27 (2000).

5. Private communication, D. Wineland.

6. Private communication, R. Blatt. 

7. Bell, J.S., On the Einstein-Podolsky-Rosen paradox, {\it Physics}, 
{\bf 1} 195-200 (1964)

8. Wigner,E.P. , On hidden variables and quantum mechanical
 probabilities {\it Am. J. Phys.}, {\bf 38}, 1005-1009 (1970).

9. Bohm, D., {\it Quantum Theory} (Prentice-Hall, Englewood Cliffs, N.J.,
1951).

10. Aspect,A. Dalibard,J. \& Roger,G., Experimental test of Bell
inequalities using time-varying analyzers, {\it Phys. Rev. Lett.},
{\bf 49}, 1804-1807, (1982).

11. Weihs,G. et al, Violation of Bell's inequality under strict locality
conditions, {\it Phys. Rev. Lett.}, {\bf 81}, 5039-5043 (1998).
 
12. Tittel,W. et al, Long-distance Bell-type tests using energy-time
entangled photons, {\it Phys. Rev. A} {\bf 59},4150-4163 (1999).

13. Percival,I.C., Quantum transfer functions, weak nonlocality and
 relativity , {\it Phys. Lett. A}, {\bf 244}, 495-501, (1998).

14. Einstein,A. Podolsky,B. \&Rosen,N., Can quantum-mechanical
description of reality be considered complete?, {\it Phys. Rev.} {\bf
47}, 777-780 (1935).

15. Santos,E., Critical analysis of the empirical tests of local
hidden-variable theories, {\it Phys. Rev. A} {\bf 46}, 3646-3656,
(1992).

16. Clauser, J.F. and Horne,M.A. {\it Phys. Rev. D} {\bf 10},
526-535, (1974).

17. Gisin,N. \& Gisin,B. Local hidden variable model of quantum
correlation exploiting the detection loophole, {\it Phys. Lett. A}
{\bf 260}, 323-327, (1999).

18. Pais,A. {\it Subtle is the Lord \dots}, (Oxford University Press, 1982) p382.

\vs{\bf Acknowledgements}

I thank Roberto Basoalto, Ed Fry, Nicolas Gisin, Serge Haroche, Thomas
Walther and Ting Yu helpful and stimulating communications and
Leverhulme, the ESF and PPARC for support.

 \end{document}